\pdfoutput=1
\documentclass[fleqn,usenatbib,usedcolumn]{mnras}
\usepackage[british]{babel}    
\usepackage{graphicx}
\usepackage{amsmath}
\usepackage{xspace}
\usepackage{ulem}
\usepackage{color}

\newcommand{\p}{\,.} 
\newcommand{\vg}{\,,} 

\newcommand\Eq[1]{Eq.~(\ref{#1})}
\newcommand\Fig[1]{Fig.~\ref{#1}}

\newcommand\Sec[1]{sec. \ref{#1}}

\newcommand{\Msun}{\mathrm{M}_{\sun}}

\newcommand\ML{M}

\newcommand\DS{D_S}
\newcommand\DL{D_L}
\newcommand\FS{F_S}

\newcommand{\tE}{t_\mathrm{E}}
\renewcommand{\to}{t_0}
\newcommand{\uo}{u_0}
\newcommand\pirel{\pi_\mathrm{rel}}
\newcommand\murel{\mu_\mathrm{rel}}
\newcommand\murelv{{\vec\mu}_\mathrm{rel}}

\newcommand\piE{\pi_\mathrm{E}}

\newcommand\piEpe{\pi_{\mathrm{E},\perp}}
\newcommand\piEpa{\pi_{\mathrm{E},\parallel}}

\renewcommand\vec[1]{\boldsymbol{#1}}

\newcommand{\mas}{\mathrm{mas}}

\newcommand{\metal}{\mbox{[Fe/H]}}

\newcommand{\Vc}{\mathcal{V}} 
\newcommand{\VcE}{\Vc_\mathrm{E}} 
\newcommand{\Vm}{V} 
\newcommand{\thE}{\theta_\mathrm{E}} 
\newcommand{\thEv}{\vec\theta_\mathrm{E}} 
\newcommand{\piEv}{\vec\pi_\mathrm{E}} 
\newcommand{\thEi}{\thE^{-1}} 
\newcommand\TF[1]{\mathrm{FT}\left[#1\right]}
\newcommand{\lbdo}{\lambda_0} 
\newcommand{\thEE}{\theta_{\mathrm{E},E}}
\newcommand{\thEN}{\theta_{\mathrm{E},N}}
\newcommand{\thEpe}{\theta_{\mathrm{E},\perp}}
\newcommand{\thEpa}{\theta_{\mathrm{E},\parallel}}

\usepackage[T1]{fontenc}
\usepackage{ae,aecompl}

\usepackage{mathptmx}
\usepackage{txfonts}

\title[Interferometric observation of microlensing events]{Interferometric observation of microlensing events}

\author[A. Cassan \& C. Ranc]{Arnaud Cassan\thanks{E-mail: \href{mailto:cassan@iap.fr}{cassan@iap.fr}} and Cl\'ement Ranc\thanks{E-mail: \href{mailto:ranc@iap.fr}{ranc@iap.fr}}
\\
Sorbonne Universit\'es, UPMC Univ Paris 6 et CNRS, UMR 7095, Institut d'Astrophysique de Paris, 98 bis bd Arago, 75014 Paris, France}

\date{Accepted 2016 February 12. Received 2016 February 11; in original form 2016 January 7}

\pubyear{2016}

\begin{document}
\label{firstpage}
\pagerange{\pageref{firstpage}--\pageref{lastpage}}
\maketitle

\begin{abstract}
Interferometric observations of microlensing events have the potential to provide unique constraints on the physical properties of the lensing systems. In this work, we first present a formalism that closely combines interferometric and microlensing observable quantities, which lead us to define an original microlensing $(u,v)$ plane. We run simulations of long-baseline interferometric observations and photometric light curves to decide which observational strategy is required to obtain a precise measurement on vector Einstein radius.  We finally perform a detailed analysis of the expected number of targets in the light of new microlensing surveys (2011+) which currently deliver 2000 alerts/year. We find that a few events are already at reach of long baseline interferometers (CHARA, VLTI), and a rate of about $6$ events/year is expected with a limiting magnitude of $K\simeq 10$. This number would increase by an order of magnitude by raising it to $K\simeq 11$. We thus expect that a new route for characterizing microlensing events will be opened by the upcoming generations of interferometers.
\end{abstract}

\begin{keywords}
gravitational lensing: micro -- techniques: interferometric -- planets and satellites: detection.
\end{keywords}



\section{Introduction}

	Gravitational microlensing was first proposed by \cite{Paczynski1986} as an observational technique to probe the dark mass content  of the Galaxy's halo. \cite{MaoPaczynski1991} further extended microlensing applications to the detection of brown dwarfs and exoplanets located in the galactic disk or bulge. Microlensing observations find today that exoplanets are ubiquitous in our Milky Way \citep{Moai}, and that free-floating exoplanets may be common as well \citep{SumiUnbound2011}. Gravitational microlensing results in the bending of light rays emitted by a background source star when they pass close to an intervening lensing massive object, such as a star or a planetary system, thereby spliting the source's disk into several images. While the typical angular separation of these images (of order of a milliarcsecond, or mas) is far too small to be resolved by classical telescopes, long-baseline interferometers of $100\,\mbox{m}$ or more can in theory resolve them. Such observations have great potential to put constraints on the mass and distance of the mirolensing system.
	
	\cite{Delplancke2001} derived the fringe visibility produced by the two point-like images of a point source lensed by a single lens, and discussed the possibility of observing them with the ESO Very Large Telescope Interferometer (VLTI). \cite{DalalLane2003} further extended this study to closure phase measurements, introduced the Einstein ring radius $\thE$ in the formalism, and performed a first estimation of the number of potential targets. Since high magnification events are the most promising targets, \cite{Rattenbury2006} studied the effect in visibility and closure phase of the spatial extension of the source for a single lens, which are then non negligible. 
	
	In this work, we first establish a new formalism that puts together interferometric and microlensing quantities, which lead us to define a microlensing $(u,v)$ plane (\Sec{sec:interomlensing}). We then discuss microlensing interferometric observables together with light curve modeling and physical parameter measurements (\Sec{sec:mlensingmodels}). In \Sec{sec:simus}, we run simulations of microlensing events observed photometrically and through interferometry to design an efficient observational strategy. We finally perform a detailed analysis of expected number of targets in the light of new generation of microlensing alert networks (\Sec{sec:observations}).

\section{Interferometric microlensing} \label{sec:interomlensing}

\subsection{Einstein ring radius} \label{sec:THEV}

	During a microlensing event, the multiple images of the source have typical separations of about the diameter of the Einstein ring whose angular radius is
\begin{equation}
	\thE=\sqrt{\kappa \ML \pirel}\vg
\label{eq:thE}
\end{equation}
while their exact position in the plane of the sky at time $t$ is given by the lens equation (\Sec{sec:models}). In \Eq{eq:thE}, $M$ is the total mass of the lens, $\pirel/ \mbox{AU}=\DL^{-1}-\DS^{-1}$ is the relative lens-source parallax (respectively located at distances $\DL$ and $\DS$ from the Sun) expressed in astronomical units (AU), and $\kappa \simeq 8.144\,\mbox{$\mas/\Msun$}$ is a constant. For standard microlensing scenarii ($M\sim 0.5 - 1\,\Msun$, $\pirel\sim 0.03-0.4\,\mas$), $\thE\sim 0.3-1.7\,\mas$; long-baseline interferometers are therefore the instruments of choice for resolving the individual images. 

	Interferometers not only have the ability to measure the angular separation of individual images, but also they measure their position in the plane of the sky. This situation is very similar to astrometric microlensing, in which the shift of the images light centroid is measured while the individual images are not resolved \citep[e.g.][]{Dominik2000}: the centroid shift is directly proportional to $\thE$, while its direction is directly linked to the lens-source relative angular motion $\murelv$ (in the observer's frame) through the microlensing model. This led \citet{GY14} to introduce a new quantity, the \textit{vector Einstein radius} (two-dimensional in the plane of the sky),
\begin{equation}
	\thEv \equiv \frac{\murelv}{\murel}\,\thE \vg
\end{equation}
whose direction is that of $\murelv$ and whose magnitude is $\thE$. This brings very interesting properties to measure the lens physical parameters when combined with other two-dimensional measurements, such as parallax (\citet{GY14}, cf. \Sec{sec:physical}). Hence we can generalize this approach to any kind of measurement, as long as it delivers an angle and a direction in the sky. Following this idea, we develop below a formalism exploiting the properties of vector $\thEv$ for interferometric observations.

\subsection{The microlensing $(u,v)$ plane} \label{sec:uvplane}

\begin{figure}
\begin{center}
	\includegraphics[width=\columnwidth]{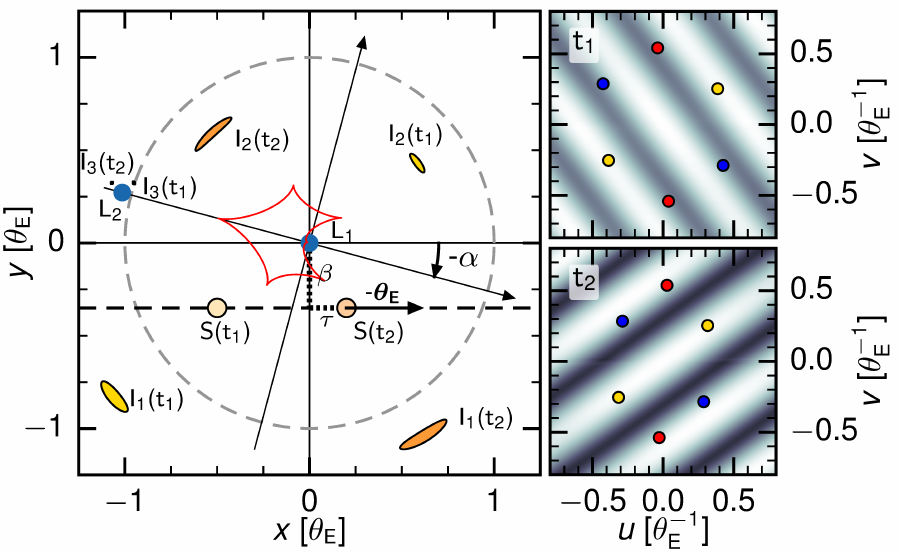}
	\caption{\textit{Left panel:} Vector Einstein radius $\thEv$ is along the $x$-axis and has same direction as the lens-source relative motion. The primary lens component $L_1$ is fixed at the center of the coordinate systems and the secondary $L_2$ is at $-se^{-i\alpha}$. The source $S$ is located at $\tau(t)+i\beta(t)$, and is shown at two epochs $S(t_1)$ and $S(t_2)$. The corresponding images are labelled $I_k(t_1)$ and $I_k(t_2)$. \textit{Right panels:} Fringe visibilities patterns at epochs $t_1$ and $t_2$ (same observing filter), illustrating microlensing supersynthesis. The colored points mark the $(u_i,v_i)$ measurements obtained with three different baselines.}
\label{fig:visibility}
\end{center}
\end{figure}

	Microlensing of a source star results in a distribution of light $I(\vec\theta)$ in the plane of the sky, where $\vec\theta=(\theta_x,\theta_y)$ is the angular position vector in physical units relative to a given $(O,x,y)$ orthonormal frame. The interferometer  measures the squared modulus of the fringe visibility, $\Vm^2=|\Vc|^2$, where $\Vc\in\mathbb{C}$ is computed \textit{via} the van Cittert-Zernike theorem,
\begin{equation}
\label{eq:vis}
	\Vc\left(\frac{B_x}{\lbdo},\frac{B_y}{\lbdo}\right) = \frac{\iint\displaystyle I(\theta_x,\theta_y)\,e^{-i2\pi\frac{\vec B\cdot\vec\theta}{\lbdo}} d\theta_x d\theta_y}{\iint I(\theta_x,\theta_y)\, d\theta_x d\theta_y} \p
\end{equation}
Here, $\vec B$ is the interferometer baseline (vector linking two telescopes) projected onto the plane of the sky, and $\lbdo$ is the wavelength of the observation. In Fourier formalism, we  equivalently write the integrals in \Eq{eq:vis} as
\begin{equation}
\label{eq:defFT}
	\TF{I}(u,v) = \iint I(x,y)\,e^{-i2\pi (ux + vy)} dxdy 
\end{equation}
with $I(x,y) = I(\vec\theta)$ and using the definition of the Fourier transform, which, from Eqs. (\ref{eq:vis}) and (\ref{eq:defFT}) implies
\begin{equation}
\label{eq:defk}
	ux + vy \equiv \frac{\vec B}{\lbdo}\cdot\vec\theta  \equiv -\vec k\cdot\vec\theta \p
\end{equation}
In this expression, we have introduced the vector $-\vec k(\lbdo, t)$, the two-dimensional projection onto the plane of the sky of $\vec B/\lbdo$ at observation time $t$. At this stage, we have not defined yet the coordinate system $(x,y)$. To make a natural link between microlensing and interferometry formalisms, we choose $(x,y)$ to be the classical microlensing coordinates of the images in the lens plane, expressed in $\thE$ units. It results from \Eq{eq:defk} that $(u,v)$ are spatial frequencies in $\thEi$ units, which is recalled by  the subscript `E' in our adopted expression (equivalent to \Eq{eq:vis}) of the fringe visibility,
\begin{equation}
\label{eq:visuv}
	\VcE(u,v) = \frac{\TF{I}(u,v)}{\TF{I}(0,0)} \p
\end{equation}
These $(u,v)$ coordinates hence define a new microlensing interferometric plane that we will call the \textit{microlensing}, or \textit{Einstein $(u,v)$ plane}. The connection between microlensing image positions and corresponding fringe visibility patterns in the Einstein $(u,v)$ plane is illustrated in the right panels of \Fig{fig:visibility}.

	It finally remains to define the orientation of the $(x,y)$ coordinate system in the plane of the sky (which also defines the orientation of $(u,v)$ since they are conjugate Fourier variables). A natural choice is to take the $x$-axis to be along $-\thEv$ with $(x,y)$ right-handed to follow usual microlensing conventions (left panel of \Fig{fig:visibility}). While $\VcE$ can be computed in the Einstein $(u,v)$ plane directly from the images position calculated from to the microlensing model (cf. \Sec{sec:models}), the actual $(u,v)$ probed by a specific measurement is a combination of the magnitude of the two components of $\thEv$ (microlensing side) and the two components of $\vec k$ (interferometry side), and reads
\begin{equation}
\label{eq:uv}
	\begin{pmatrix}
		u \\
		v 
	\end{pmatrix}
	\equiv 
	\begin{pmatrix}
		\thEv\cdot\vec k \\
		\thEv\times \vec k
	\end{pmatrix}
	= 
	\begin{pmatrix}
		\thEN k_N + \thEE k_E \\
		\thEN k_E - \thEE k_N
	\end{pmatrix}
	= 
	\begin{pmatrix}
		\thEpa k_\parallel + \thEpe k_\perp \\
		\thEpa k_\perp - \thEpe k_\parallel
	\end{pmatrix}\p
\end{equation}
In the two latter expressions, $\thEv$ has been first decomposed in the North-East frame ($N, E$), while in the second case it has been decomposed in a parallel-perpendicular frame ($\parallel, \perp$) related to parallax measurements. These aspects are detailed in \Sec{sec:physical}.

\subsection{Microlensing supersynthesis} \label{sec:mlsupersyn}

	Interferometric observations consist in sampling the $(u,v)$ plane at given epochs $t_i$ and measure the corresponding fringe visibilities at microlensing spacial frequencies $(u_i,v_i)$. 
	
	All possible combinations of 2 telescopes amongst $N$ will give rise to $N!/2\,(N-2)!$ possible baselines, and the same number of pairs of $(u_i,v_i)$ data points. When three  (or more) telescopes are available, it is possible to build the complex product of the individual complex visibilities and measure the so-called closure phase \citep[e.g.][]{DalalLane2003,Rattenbury2006}, ${\phi_\mathrm{E}}_{,123} = \arg\left({\VcE}_{,12}{\VcE}_{,23}{\VcE}_{,31}\right)$.
In fact, a right arrangement of three baselines gives $\vec B_{12}+\vec B_{23}+\vec B_{31}=\vec 0$, which implies that the phase errors (due to atmospheric turbulence) in the visibility of the individual baselines cancel out, resulting in a well-measured \textit{microlensing closure phase}. Measuring this quantity is particularly interesting because the expected signal-to-noise ratio is lower than that of the visibility \citep{DalalLane2003}. In practice however, closure phase works well with three telescopes but is challenging with more telescopes . 
		
	Once the observing baselines are chosen, the simplest way to further sample the $(u,v)$ plane is to use the rotation of the Earth, a technique called supersynthesis: $u$ and $v$ map the  $(u,v)$ plane as $\vec k$ varies time according to \Eq{eq:uv}. The same equation shows that changing the wavelength $\lbdo$ of observation  also affects $\vec k$, and additional data points in the $(u,v)$ plane are obtained when multi-band observations are performed (e.g. in $H$ and $K$). 

	Finally, microlensing itself provides an intrinsic \textit{microlensing supersynthesis}: as the source moves relative to the lens, the microlensed images will change in position and shape, resulting in a change of the visibility pattern. Depending on the configuration, this change can range from barely noticeable to very strong. In the case of a single lens for example, the two diametrically opposite images rotate with time around the Einstein ring, and one can show that their maximum rotation rate $\omega$ (rad/s) is given by $\omega \simeq 1/\uo\tE$, where $\uo\ll 1$ is the closest approach between the source and lens in $\thE$ units and $\tE$ the event's timescale. 

\section{Microlensing models} \label{sec:mlensingmodels}

\subsection{Point-source single and binary lenses} \label{sec:models}

	The multiple image positions of a point-source lensed by a binary-mass object with mass ratio $q<1$ are given by the complex lens equation \citep{Witt1990}, 
\begin{equation}
\label{eq:lenseq}
	\zeta = z - \frac{1}{1+q}\left(\frac{1}{\overline{z}} + \frac{q}{\overline{z}+se^{i\alpha}} \right) \vg
\end{equation}
where $\zeta=\xi+i\eta$ is the affix of the source and $z=x+iy$ is the affix of the images $z_k$ found by solving the lens equation (for a binary lens, there are three or five solutions for $z$, and only two for a single lens). Both $\zeta$ and $z$ are in $\thE$ units. Following the convention of \citet{causfix}, the primary body is at the center of the coordinate system, and the source trajectory makes an angle $\alpha$ with respect to the binary lens axis (with the secondary on the left). It results from the choice made in \Sec{sec:uvplane} ($x$-axis along $\thEv$) that the affix of the secondary lens is $-se^{-i\alpha}$, where $s>0$ is the binary-lens separation. The single lens equation is obtained setting $q=0$. 

	Each point-like image $k$ has a flux magnified by a factor $\mu_k = |\partial\zeta/\partial z|^{-1}$ with respect to the source flux $\FS$, and thus contributes an additive term $\mu_k\FS\delta(x-x_k, y-y_k)$ to the total intensity $I(x, y)$. The corresponding complex visibility (in $\thEi$ units) then reads
\begin{equation}
	\label{eq:VPS}
	\VcE(u,v) = \frac{1}{\sum_k\mu_k} \sum_k \mu_k e^{-i2\pi (u x_k + v y_k)} \p
\end{equation}
Contrasts are higher when at least two of the $\mu_k$ are close in magnitude.

\subsection{Finite-source effects}  \label{sec:finite}

	The effect of the finite size of the source has been studied in detail by \citet{Rattenbury2006} in the single lens case. Images are then spatially extended and distorted, and form  \textit{macro images}. For example, a ring-like image is a merger of two extended, single-lens images. The authors found that finite-source effects become significant at high magnification, when the two images are very elongated along the Einstein ring. We can generalize this finding to the case of  binary lenses: when the source crosses a caustic, it generates a macro image which has an elongated shape, and which angle relative to the critical line and ellipticity can be evaluated through a Taylor expansion of the lens equation \citep[like that of][]{SchneiderWeiss1986}. 
	
	To compute the visibility from \Eq{eq:visuv}, numerical integration is required since, obviously, there exists no analytical formula \citep[even in the single lens case,][]{Rattenbury2006}. Contouring and inverse ray shooting methods provide integration methods of choice, by analogy with magnification. Contouring methods \citep{Bozza2010,Dominik2007cont} first calculate the macro images contours of the extended source, which in practice generates many difficulties \citep[e.g.][]{Dong2006}. Once the oriented contours $\partial \mathcal{I} =(X,Y)$ are drawn for each macro image, Green-Riemann's formula provides an easy and inexpensive way to compute the visibility by replacing surface integrals in \Eq{eq:defFT} by contour integrals, such as
\begin{equation}
	- \frac{i}{2\pi} \oint_{\partial\mathcal{I}} \frac{e^{-i2\pi (uX + vY)}}{v} dX = \frac{i}{2\pi} \oint_{\partial\mathcal{I}} \frac{e^{-i2\pi (uX + vY)}}{u} dY 
\end{equation}
in the case of a uniformly bright source with $I(x,y)=1$. Limb-darkened sources are treated as nested uniform disks forming a number of annulus of same intensity. Inverse ray shooting \citep{Wambsganss1997} can be used to compute the visibility, provided that each ray carries complex factor $e^{-i2\pi (ux + vy)}$ of  the pixel point $x+iy$ in the lens plane it was shot from. Improvements of these methods using  contouring and ray shooting of a larger source  \citep{Dong2006} can be easily adapted to visibility calculations.

\subsection{Blend sources}  \label{sec:blend}

	Microlenses are compact massive objects located at $\DL \sim 1-8$ kpc from Earth, but only stars are bright enough to introduce a significant blend contribution to the total light. At these distances, typical angular radius of stars range from $0.1-10\,\mu\mbox{as}$, and are not resolved by the interferometer. Hence the lens, as well as other blending sources $l$, contribute an additive term $g_l\FS\delta(x-x_l, y-y_l)$ to $I(x,y)$, with $g_l$ the corresponding blend ratio relative to $\FS$ in the observing passband. In the visibility formula \Eq{eq:VPS}, $g_l  e^{-i2\pi (u x_l + v y_l)}$ terms further enter the sum while the normalization is changed to the sum of all $\mu_k$ and $g_l$.
	
	Blending sources decrease the global contrast of the visibility, and should be included in the calculation (although this aspect has not been considered in previous studies). In general, $g_l$ can be estimated with enough precision from the photometric monitoring. More interestingly, if a point-source blend is bright enough, in principle its intensity can be measured directly from the interferometric observation, while this is usually achieved only via high resolution imaging.

\subsection{Model parameters and lens physical parameters} \label{sec:physical}

	With the lensing system fixed in the reference frame, the source trajectory is usually modeled through two time-dependent quantities,
\begin{equation}
\label{eq:taubeta}
	\tau = \frac{t-\to}{\tE} + \delta\tau  \vg \quad \beta = \uo + \delta\beta \vg
\end{equation}
where $\tau$ is along the source motion and $\beta$ in the perpendicular direction, as seen in \Fig{fig:visibility}; $\uo$ is the minimum distance between the source and the lens primary, $\to$ is the corresponding date, and $\tE$ is the time it takes for the source to travel one $\thE$. The two  correction terms $\delta\tau$ and $\delta\beta$ are non-trivial and time-dependent when parallax effects are significant \citep{Gould1994}.

	The light curve model provides the parameters of the lens and trajectory such as $q$, $s$, $\tE$, $\alpha$, $\to$ and $\uo$ aforementioned, but in favorable cases also a measurement of the parallax vector $\piEv$ or the source size $\rho$ in $\thE$ units. The best photometric model then predicts the shape and position of the images in $\thE$ units at any time $t$ (with a given uncertainty) and thus yields the corresponding visibility pattern in $\thEi$ units at $t$, which can be compared to interferometric data points in the Einstein $(u,v)$ plane (right panels of \Fig{fig:visibility}). Then, the two components of $\thEv$ are adjusted as two independent parameters to match the values of the predicted and measured Einstein $(u_i,v_i)$ through Bayesian (MCMC, DEMC) algorithms \citep[e.g.][]{Kains2012,UCausfix}. The advantage of the formalism developed here is that two components of $\thEv$ in \Eq{eq:uv} are constrained separately with potentially different probability distribution widths.
	
	As seen in \Eq{eq:thE}, $\thE$ is a combination of the lens mass $\ML$ and distance $\DL$ through $\pirel$, since $\DS$ is usually well-known from color-magnitude diagrams (in most cases, the source is in the Galactic bulge, at $\DS\simeq 7.6\,\mbox{kpc}$). Several second order effects can be used to constrain these parameters, and can even lead an over-constrained problem \citep{Ranc2015}. In particular, when $\thEv$ is measured, quantities
\begin{equation}
	 \murelv = \thEv/\tE \vg \quad \vec{v}_\perp = \DL\,\murelv \vg
\label{eq:vperp}
\end{equation}	
are immediately found, since $\tE$ is measured from the light curve fit. Here, $\vec{v}_\perp$ is the physical lens-source speed (km/s) at the lens position, which can be directly compared to predictions of Galactic models. The measure of $\thEv$ also provides an independent lens mass-distance relation,
\begin{equation}
\label{eq:massdist}
	\ML\pirel = \thE^2/\kappa \p
\end{equation}
Combined with the measurement of vector parallax $\piEv$ (which has same direction as $\thEv$ and amplitude $\piE = \pirel/\thE$), \Eq{eq:massdist} directly yields the lens mass,
\begin{equation}
	\ML = \frac{\thE}{\kappa\piE} = \frac{\thEpa}{\kappa\piEpa}\p
\label{eq:MpiE}
\end{equation}
In the first case, the modulus of $\thE$ and $\piE$ are used, but much more precise measurements can be obtained if individual components of these quantities are used (last term). In fact, as argued by \citet{GY14} in the case of astrometric microlensing, $\piEpa$ is much better constrained than $\piEpe$ or $\piE$, because $\piEpa$ undergoes a much larger variation than $\piEpe$. The lens distance is finally obtained through \Eq{eq:massdist},
\begin{equation}
\label{eq:pirelval}
	\pirel = \thE^2\frac{\piEpa}{\thEpa} \p
\end{equation}

\section{Observational strategy} \label{sec:simus}

\begin{figure}
\begin{center}
	\includegraphics[width=\columnwidth]{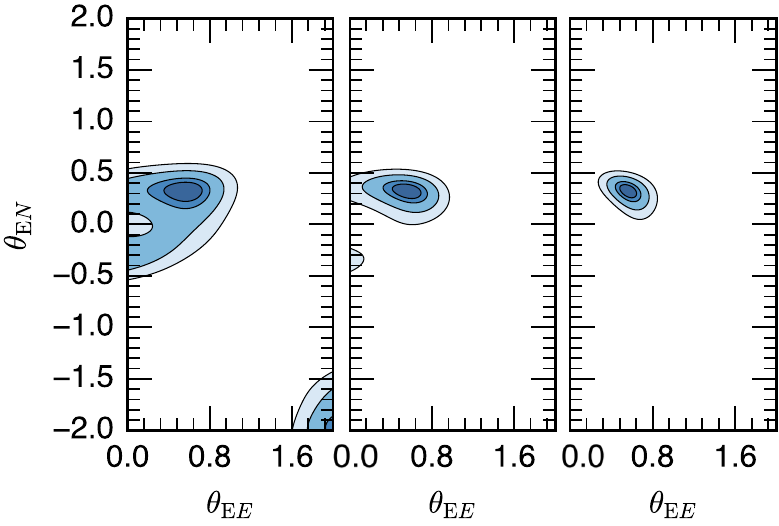}
\caption{Confidence contours (1 to 4 $\sigma$) on the North (vertical axis) and East (horizontal axis) components of $\vec{\thE}$ with one (left), two (middle) or three (right) simulated observations with VLTI/PIONIER. Good constraints are obtained with two or more observations. }
\label{fig:thE}
\end{center}
\end{figure}

	The goal of the interferometric observation is to measure the two components of vector Einstein radius $\vec{\thE}$.  In terms of observational strategy, at a given time, more than a hundred microlensing events are in progress. The most promising events are followed-up at high photometric cadence by  survey telescopes OGLE\footnote{\url{http://www.astrouw.edu.pl/$\sim$ogle}} (Optical Gravitational Lensing Experiment), MOA\footnote{\url{http://www.phys.canterbury.ac.nz/moa}} (Microlensing Observations in Astrophysics), KMTnet (Korea Microlensing Telescope Network)\footnote{\url{http://www.kasi.re.kr/english/Project/KMTnet.aspx}} and monitored by networks of telescopes such as RoboNet\footnote{\url{http://robonet.lcogt.net}} (Las Cumbres Observatory LCOGT), PLANET\footnote{\url{http://planet.iap.fr}} (Probing Lensing Anomalies NETwork),  $\mu$FUN\footnote{\url{http://www.astronomy.ohio-state.edu/$\sim$microfun}} or MiNDSTEp\footnote{\url{http://www.mindstep-science.org}}. The  interferometric targets have to be identified in this large number of ongoing events. One of the main difficulty is to predict the peak magnitude of the events in advance, but experience of real-time modelling shows that a fair estimation can usually be obtained two (sometimes three) days in advance, which is just good for issuing a Target of Opportunity observation 48h before the peak. We discuss the criteria to choose the targets in \Sec{sec:observations}. 
	
	To illustrate how interferometric observations must be performed for a good constraint on  considered a typical single lens characterized by $\uo=0.01$, $\tE=30\,\mathrm{d}$ and an Einstein radius with North-East components $\thEN=0.325\,\mas$ and $\thEE=0.563\,\mas$ ($\thE=0.650\,\mas$). To fix ideas, we simulate an observation with the ESO/VLTI PIONIER instrument, which combines the light from 4 Auxiliary Telescopes (ATs) at a time, leading to six simultaneous baselines. We take into account the errors on the visibility by adding a gaussian noise to the squared visibilities of 3\%. In practice, at each observation epoch, we compute $\vec{k}$ (cf. \Eq{eq:defk}) from the Galactic coordinates of the event and the location of the telescopes. The time dependence of this quantity is due to the Earth's rotation (supersynthesis), while the visibility pattern depends on the particular geometry of the images for the given  $\tau$ and $\beta$ at the time of the observation (microlensing supersynthesis). The two components of $\vec{\thE}$ are then free parameters (cf. \Eq{eq:uv}) that need to be fitted. For each probed $(\thEN, \thEE)$, we compute the corresponding spatial frequencies in the Einstein $(u,v)$ plane through \Eq{eq:uv} and the final squared visibility with \Eq{eq:vis}. 
	
	The resulting confidence intervals on $\vec\thE$ (1 to 4 $\sigma$) obtained after one, two and three observations around the peak of the event are drawn in \Fig{fig:thE}. A seen in the figure, very good constraints are obtained with two (or more) measurements. Hence, a good observing strategy is that each microlensing event be observed at least two times, but an additional third observation is a plus to ensure a good measurement at magnitudes close to the sensitivity limits of the instruments. Furthermore, observations at three epochs close in time (spread over about 48h) should show up the displacement with time of the multiple images.

\section{Interferometric microlensing targets} \label{sec:observations}

	New generations of alert telescopes (2011+) have increased the rate of microlensing event detections from a few hundreds to more than 2000 per year. This provides an unprecedented ground for predicting interferometric microlensing targets. Here we use data of all microlensing events alerted by the OGLE-IV Early Warning System\footnote{EWS: \url{http://ogle.astrouw.edu.pl/ogle4/ews/ews.html}} \citep{UdalskiOGLEIV} during seasons 2011-14 (4 years, $\sim 7000$ events) to perform a precise estimation of a mean number of targets per year as a function of interferometer $K$-limit magnitude. 
	
\begin{figure}
\begin{center}
	\includegraphics[width=\columnwidth]{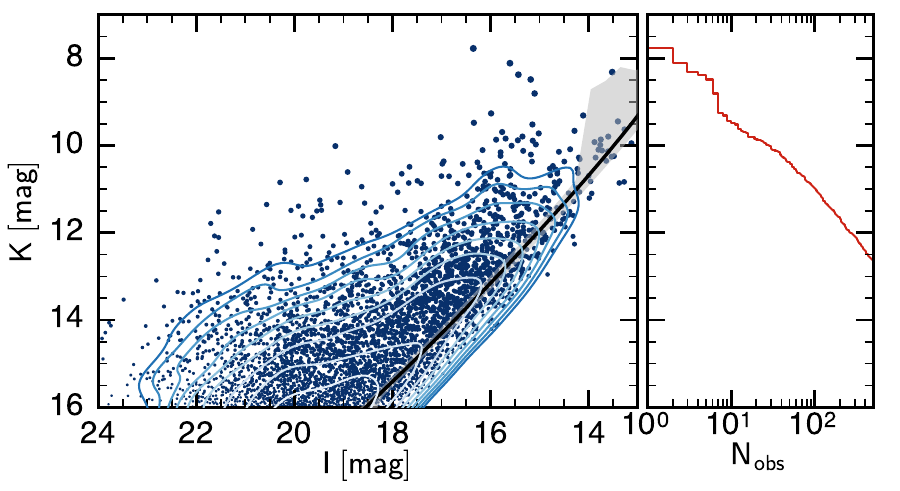}
\caption{\textit{Left panel:} Predicted instrumental $K$ at peak \textit{vs.} de-blended $I$ of the source for microlensing events alerted by OGLE in 2011-14 which passed our selection criteria (blue dots). The dark gray line is a reference isochrone of age $8\,\mbox{Gyr}$ and metallicity $\metal=-0.2$ for a source at $7.6\,\mbox{kpc}$, and is used to derive the source's $K$ magnitude after treatment of extinction. The light gray shading indicates the typical dispersion around the reference isochrone in all these parameters. Blue contour lines (logarithmic levels) draw the probability density of the dots distribution. \textit{Right panel:} Cumulative histogram of the number of events that have a $K$ peak magnitude lower than $K$.}
\label{fig:statobs}
\end{center}
\end{figure}

	To do so, for every individual OGLE event we first correct from extinction the source (de-blended) $I$ baseline magnitude using $A_I$ maps from \citet{estimAI}. These de-reddened $I$ magnitudes are compared to a reference isochrone \citep{Girardi2002} of age $8\,\mbox{Gyr}$ and metallicity $\metal=-0.2$, and assuming the source is located at $7.6\,\mbox{kpc}$. This isochrone is displayed as the dark gray thick line in the left panel of \Fig{fig:statobs}, while the light gray shaded area indicates dispersion around this central value for isochrones spanning ages, metallicities and source distances of respectively $5$ to $10$ Gyr, $-0.5$ to $0.2$ dex and $6$ to $8\,\mbox{kpc}$. From this we derive magnitudes in $K$, which are then corrected from microlensing maximum magnification ($-2.5\log A$) and reddened using $A_K$ maps from  \citet{estimAK}, which  yields the predicted instrumental $K$ magnitudes at peak.
	
	At this point, we systematically remove all events that have $\delta = I_\mathrm{p}-I_\mathrm{m} \leq - 0.8$, where $I_\mathrm{p}$ is the magnitude at peak and $I_\mathrm{m}$ the brightest magnitude measured. This criterion is aimed at removing events for which the brightest data are well below the model-predicted magnitude at peak. It proves very efficient to detect events with badly covered peaks, which results in unrealistically high predicted magnifications. Cases where $\delta \geq 0$ are all kept in the final sample, since they appear to be almost always binary-lens events with minimum magnitude underestimated by single-lens models. These criteria are conservative in the sense they tend to underestimate the number of favorable events. When $-0.8<\delta<0$, we examine individually all events with $K\leq 10$ and select 26 events out of 27 selected by previous criteria. The final sample is shown as blue dots in \Fig{fig:statobs}. Blue contours draw  logarithmic levels of a non-parametric estimation of the probability density of the resulting distribution. 
	
	In the right panel of \Fig{fig:statobs}, we show the cumulative histogram of the number of events with $K$ peak magnitudes lower than $K$. The first potential target appears at $K\simeq 7.8$, while 26 events already have $K\leq 10$ (hence, a mean of $6-7$/per year). The CHARA interferometer (Center for High Resolution Astronomy) has limiting magnitudes of $K\simeq 9$, but  can reach $10$ in exceptional cases. These magnitudes are also at reach of VLTI using not only Unit Telescopes (UT, 8m), but also Auxiliary Telescopes (AT, 1.8m) as we used in the simulations presented in the previous section. From our study, an increase of only one magnitude would already provide an order of magnitude more microlensing targets for the next generation of instruments. 
	
	Although these numbers are based on observed microlensing light curves (including data loss due to bad weather for example), in practice, specific technical constraints may lessen the actual number of available targets. Such a constraint is for example the availability of a suitable guide star. Microlensing events are always observed towards crowded fields in the direction of the galactic bulge, where ``bright stars'' ($V \sim 13$) can usually be found within $\sim 0.5-1$ arcmin of the target, so only a fraction of event should be affected.

\section{Conclusion and perspectives}
	
	New perspectives of interferometric observations of microlensing events have been opened by recent improvements in the sensitivity of long baseline interferometers such as CHARA and VLTI, and we have shown  that several microlensing events per year are already at reach. The observational strategy requires a rapid-response microlensing photometric follow-up and efficient alert system, which are already in place. Interferometric microlensing observations carry great promises to characterize completely many more microlensing systems in a near future. Future instruments such as ESO/GRAVITY are expected to greatly increase the number of microlensing events  monitored by interferometers in the coming years.

\section*{Acknowledgements}

The authors are grateful to the OGLE collaboration for providing online data and basic parameters of all microlensing events. The authors thank V. Foresto, S. Ridgway and the CHARA team for collaborating on testing the full observational strategy in a real case in May 2015. This work was supported by Universit\'e Pierre et Marie Curie grant \'Emergence-UPMC 2012.




\bibliographystyle{mnras}
\bibliography{bibli} 

\begin{thebibliography}{}
\makeatletter
\relax
\def\mn@urlcharsother{\let\do\@makeother \do\$\do\&\do\#\do\^\do\_\do\%\do\~}
\def\mn@doi{\begingroup\mn@urlcharsother \@ifnextchar [ {\mn@doi@}
  {\mn@doi@[]}}
\def\mn@doi@[#1]#2{\def\@tempa{#1}\ifx\@tempa\@empty \href
  {http://dx.doi.org/#2} {doi:#2}\else \href {http://dx.doi.org/#2} {#1}\fi
  \endgroup}
\def\mn@eprint#1#2{\mn@eprint@#1:#2::\@nil}
\def\mn@eprint@arXiv#1{\href {http://arxiv.org/abs/#1} {{\tt arXiv:#1}}}
\def\mn@eprint@dblp#1{\href {http://dblp.uni-trier.de/rec/bibtex/#1.xml}
  {dblp:#1}}
\def\mn@eprint@#1:#2:#3:#4\@nil{\def\@tempa {#1}\def\@tempb {#2}\def\@tempc
  {#3}\ifx \@tempc \@empty \let \@tempc \@tempb \let \@tempb \@tempa \fi \ifx
  \@tempb \@empty \def\@tempb {arXiv}\fi \@ifundefined
  {mn@eprint@\@tempb}{\@tempb:\@tempc}{\expandafter \expandafter \csname
  mn@eprint@\@tempb\endcsname \expandafter{\@tempc}}}

\bibitem[\protect\citeauthoryear{{Bozza}}{{Bozza}}{2010}]{Bozza2010}
{Bozza} V.,  2010, \mn@doi [\mnras] {10.1111/j.1365-2966.2010.17265.x}, \href
  {http://adsabs.harvard.edu/abs/2010MNRAS.408.2188B} {408, 2188}

\bibitem[\protect\citeauthoryear{{Cassan}}{{Cassan}}{2008}]{causfix}
{Cassan} A.,  2008, \aap, 491, 587

\bibitem[\protect\citeauthoryear{{Cassan}, {Horne}, {Kains}, {Tsapras}  \&
  {Browne}}{{Cassan} et~al.}{2010}]{UCausfix}
{Cassan} A.,  {Horne} K.,  {Kains} N.,  {Tsapras} Y.,   {Browne} P.,  2010,
  \mn@doi [\aap] {10.1051/0004-6361/200913755}, \href
  {http://adsabs.harvard.edu/abs/2010A%26A...515A..52C} {515, A52}

\bibitem[\protect\citeauthoryear{{Cassan} et~al.,}{{Cassan}
  et~al.}{2012}]{Moai}
{Cassan} A.,  et~al., 2012, \mn@doi [\nat] {10.1038/nature10684}, \href
  {http://adsabs.harvard.edu/abs/2012Natur.481..167C} {481, 167}

\bibitem[\protect\citeauthoryear{{Dalal} \& {Lane}}{{Dalal} \&
  {Lane}}{2003}]{DalalLane2003}
{Dalal} N.,  {Lane} B.~F.,  2003, \mn@doi [\apj] {10.1086/374549}, \href
  {http://adsabs.harvard.edu/abs/2003ApJ...589..199D} {589, 199}

\bibitem[\protect\citeauthoryear{{Delplancke}, {G{\'o}rski}  \&
  {Richichi}}{{Delplancke} et~al.}{2001}]{Delplancke2001}
{Delplancke} F.,  {G{\'o}rski} K.~M.,   {Richichi} A.,  2001, \mn@doi [\aap]
  {10.1051/0004-6361:20010783}, \href
  {http://cdsads.u-strasbg.fr/abs/2001A%26A...375..701D} {375, 701}

\bibitem[\protect\citeauthoryear{{Dominik}}{{Dominik}}{2007}]{Dominik2007cont}
{Dominik} M.,  2007, \mnras, 377, 1679

\bibitem[\protect\citeauthoryear{{Dominik} \& {Sahu}}{{Dominik} \&
  {Sahu}}{2000}]{Dominik2000}
{Dominik} M.,  {Sahu} K.~C.,  2000, \mn@doi [\apj] {10.1086/308716}, \href
  {http://cdsads.u-strasbg.fr/abs/2000ApJ...534..213D} {534, 213}

\bibitem[\protect\citeauthoryear{{Dong} et~al.,}{{Dong}
  et~al.}{2006}]{Dong2006}
{Dong} S.,  et~al., 2006, \apj, 642, 842

\bibitem[\protect\citeauthoryear{{Girardi}, {Bertelli}, {Bressan}, {Chiosi},
  {Groenewegen}, {Marigo}, {Salasnich}  \& {Weiss}}{{Girardi}
  et~al.}{2002}]{Girardi2002}
{Girardi} L.,  {Bertelli} G.,  {Bressan} A.,  {Chiosi} C.,  {Groenewegen}
  M.~A.~T.,  {Marigo} P.,  {Salasnich} B.,   {Weiss} A.,  2002, \aap, 391, 195

\bibitem[\protect\citeauthoryear{{Gould}}{{Gould}}{1994}]{Gould1994}
{Gould} A.,  1994, \mn@doi [\apjl] {10.1086/187190}, \href
  {http://adsabs.harvard.edu/abs/1994ApJ...421L..71G} {421, L71}

\bibitem[\protect\citeauthoryear{{Gould} \& {Yee}}{{Gould} \&
  {Yee}}{2014}]{GY14}
{Gould} A.,  {Yee} J.~C.,  2014, \mn@doi [\apj] {10.1088/0004-637X/784/1/64},
  \href {http://adsabs.harvard.edu/abs/2014ApJ...784...64G} {784, 64}

\bibitem[\protect\citeauthoryear{{Kains}, {Browne}, {Horne}, {Hundertmark}  \&
  {Cassan}}{{Kains} et~al.}{2012}]{Kains2012}
{Kains} N.,  {Browne} P.,  {Horne} K.,  {Hundertmark} M.,   {Cassan} A.,  2012,
  \mn@doi [\mnras] {10.1111/j.1365-2966.2012.21813.x}, \href
  {http://adsabs.harvard.edu/abs/2012MNRAS.426.2228K} {426, 2228}

\bibitem[\protect\citeauthoryear{{Mao} \& {Paczynski}}{{Mao} \&
  {Paczynski}}{1991}]{MaoPaczynski1991}
{Mao} S.,  {Paczynski} B.,  1991, \apjl, 374, L37

\bibitem[\protect\citeauthoryear{{Marshall}, {Robin}, {Reyl{\'e}}, {Schultheis}
   \& {Picaud}}{{Marshall} et~al.}{2006}]{estimAK}
{Marshall} D.~J.,  {Robin} A.~C.,  {Reyl{\'e}} C.,  {Schultheis} M.,   {Picaud}
  S.,  2006, \mn@doi [\aap] {10.1051/0004-6361:20053842}, \href
  {http://cdsads.u-strasbg.fr/abs/2006A%26A...453..635M} {453, 635}

\bibitem[\protect\citeauthoryear{{Nataf} et~al.,}{{Nataf}
  et~al.}{2013}]{estimAI}
{Nataf} D.~M.,  et~al., 2013, \mn@doi [\apj] {10.1088/0004-637X/769/2/88},
  \href {http://cdsads.u-strasbg.fr/abs/2013ApJ...769...88N} {769, 88}

\bibitem[\protect\citeauthoryear{{Paczynski}}{{Paczynski}}{1986}]{Paczynski1986}
{Paczynski} B.,  1986, \apj, 304, 1

\bibitem[\protect\citeauthoryear{{Ranc} et~al.,}{{Ranc}
  et~al.}{2015}]{Ranc2015}
{Ranc} C.,  et~al., 2015, \mn@doi [\aap] {10.1051/0004-6361/201525791}, \href
  {http://adsabs.harvard.edu/abs/2015A%26A...580A.125R} {580, A125}

\bibitem[\protect\citeauthoryear{{Rattenbury} \& {Mao}}{{Rattenbury} \&
  {Mao}}{2006}]{Rattenbury2006}
{Rattenbury} N.~J.,  {Mao} S.,  2006, \mn@doi [\mnras]
  {10.1111/j.1365-2966.2005.09769.x}, \href
  {http://cdsads.u-strasbg.fr/abs/2006MNRAS.365..792R} {365, 792}

\bibitem[\protect\citeauthoryear{{Schneider} \& {Weiss}}{{Schneider} \&
  {Weiss}}{1986}]{SchneiderWeiss1986}
{Schneider} P.,  {Weiss} A.,  1986, \aap, 164, 237

\bibitem[\protect\citeauthoryear{{Sumi} et~al.,}{{Sumi}
  et~al.}{2011}]{SumiUnbound2011}
{Sumi} T.,  et~al., 2011, \mn@doi [\nat] {10.1038/nature10092}, \href
  {http://adsabs.harvard.edu/abs/2011Natur.473..349S} {473, 349}

\bibitem[\protect\citeauthoryear{{Udalski}, {Szyma{\'n}ski}  \&
  {Szyma{\'n}ski}}{{Udalski} et~al.}{2015}]{UdalskiOGLEIV}
{Udalski} A.,  {Szyma{\'n}ski} M.~K.,   {Szyma{\'n}ski} G.,  2015, \actaa,
  \href {http://adsabs.harvard.edu/abs/2015AcA....65....1U} {65, 1}

\bibitem[\protect\citeauthoryear{{Wambsganss}}{{Wambsganss}}{1997}]{Wambsganss1997}
{Wambsganss} J.,  1997, \mnras, 284, 172

\bibitem[\protect\citeauthoryear{Witt}{Witt}{1990}]{Witt1990}
Witt H.~J.,  1990, \aap, 236, 311

\makeatother
\end{thebibliography}




\bsp	
\label{lastpage}
\end{document}